# Contact Resistance Study of Various Metal Electrodes with CVD Graphene

Amit Gahoi, Stefan Wagner, Andreas Bablich, Satender Kataria,

Vikram Passi, Max C. Lemme

University of Siegen, School of Science and Technology, Hölderlinstr. 3,

57076 Siegen, Germany

Corresponding author: max.lemme@uni-siegen.de

ABSTRACT

In this study, the contact resistance of various metals to chemical vapour deposited (CVD) monolayer graphene is investigated. Transfer length method (TLM) structures with varying widths and separation between contacts have been fabricated and electrically characterized in ambient air and vacuum condition. Electrical contacts are made with five metals: gold, nickel, nickel/gold, palladium and platinum/gold. The lowest value of 92 $\Omega\mu$m is observed for the contact resistance between graphene and gold, extracted from back-gated devices at an applied back-gate bias of -40 V. Measurements carried out under vacuum show larger contact resistance values when compared with measurements carried out in ambient conditions. Post processing annealing at 450°C for 1 hour in argon-95% / hydrogen-5% atmosphere results in lowering the contact resistance value which is attributed to the enhancement of the adhesion between metal and graphene. The results presented in this work provide an overview for potential contact engineering for high performance graphene-based electronic devices.

INTRODUCTION

In order to fully exploit the exceptional properties of graphene, the electrical contact resistance ($R_C$) associated with the metal-graphene junction should be low. In fact, low values of $R_C$ is identified as a figure of merit for high frequency graphene based field effect transistors (GFET), as it translates into higher maximum oscillation frequency ($F_{max}$) and larger on-state current ($I_{ON}$) [1], [2]. The contact resistivity requirement for state-of-the-art silicon metal oxide field effect transistors (MOSFETs) is 80 Ωμm [3]. The lowest values of $R_C$ [4] reported in literature for metal-graphene contacts are close to that number. Most reported $R_C$ values, however, are higher than those for complementary metal oxide semiconductor (CMOS) technology and vary over a wide range: from few hundreds [5] to few thousands of Ωμm [6] . This variation of $R_C$ can be attributed to (i) variations in graphene growth conditions, (ii) variations in process steps of fabrication, (iii) metals used for the contacts, (iv) values of gate bias and (v) variations of measurement conditions. High $R_C$ arises due to the current injection from a 3D material (metal) to a 2D material (graphene), which typically leads to a line contact instead of an areal contact. Figure 1. shows the schematic energy-density of states relation for metal and graphene, depicting limited density of states available in graphene under the metal contact.

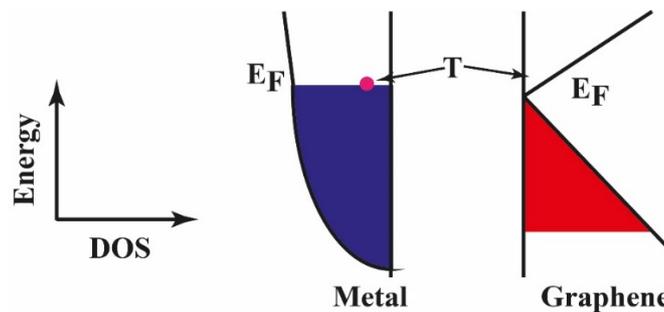

Figure 1: Schematic of the energy-density of states relation for metal and graphene, showing limited density of states available in graphene under the metal contact (after [9]).

Two important factors thus play significant roles in $R_C$: (i) the limited DOS available in the graphene under the contact and (ii) the metal-graphene coupling [5], [10], [11]. Doping of graphene to increase the DOS using conventional methods such as ion implantation is technically challenging [12] due to the thermodynamically stable carbon-carbon bonds in graphene that prevent the substitution of carbon by either boron or nitrogen [13]. Nevertheless, environmental effects such as humidity influence the DOS of monolayer graphene [9], [14]. The metal-graphene coupling leads to charge transfer at the interface as illustrated in Figure 2. The amount of charge transfer gradually decreases away from the metal-graphene junction. The transferred electrons shift the Fermi energy ($E_F$) in the graphene significantly and leads to the formation of a dipole which has a potential difference expressed as $\Delta V$ [10]. $\Delta V$ depends on the bonding between the metal and the graphene and it differs from the work function difference between graphene and metal $\Delta \Phi_{MG} = \Phi_M - \Phi_G$ where $\Phi_M$ is the work function of metal, $\Phi_G$ is the work function of graphene [10]. The interface dipole formation due to the charge transfer increases the contact resistance. Since the low DOS around the Fermi level ($E_F$) of intrinsic graphene increases $\lambda$, the charge transfer region (defined by the screening length) may give rise to a p-n junction in the graphene, which results in an additional resistive component.

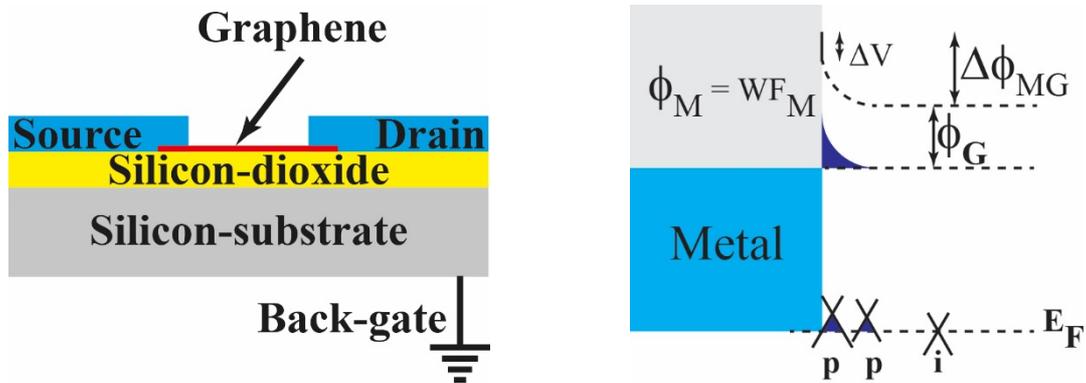

Figure 2: (a) Schematic of a back-gated FET (b) Energy band diagram of the metal-graphene contact (after [10]).

Efforts to reduce $R_C$ include metal work function engineering [15] [11] and modified contact geometry [16], [17]. Contamination at the metal-graphene interface introduced during device fabrication (graphene transfer and subsequent photolithography steps) also contributes to high $R_C$ values [18].

Here, we report a systematic and detail investigation of contact resistivity and sheet resistance extracted using transfer length method (TLM) structures at different backgate bias for various metal electrodes with CVD Graphene. We have found, for the first time, a direct correlation between sheet resistance and contact resistance which is consistently observed under all experimental conditions and contact metals used.

EXPERIMENT

Thermally oxidized (85 nm) p-Silicon <100> wafers with a boron doping concentration of $3 \times 10^{15}$ cm$^{-3}$ were used as starting substrates. The wafers were diced into samples of 1.3 cm x 1.3 cm prior to device fabrication. The samples were cleaned in an ultrasonic bath containing acetone, followed by cleaning in isopropyl alcohol (IPA) and finally rinsed with deionized water (DI water). Large area graphene was grown on copper (Cu) foil in a NanoCVD (Moorfield, UK) rapid thermal processing tool by the CVD method [19] as shown in Figure 3a. Graphene growth was performed in a three step process where (i) Cu foil is first annealed at 950°C in a hydrogen (20%) atmosphere, then, (ii) argon (80%), methane (5%) and hydrogen (15%) gas is passed into the chamber where the methane dissociates, leading to graphene growth, and finally (iii) the Cu foil is cooled to room temperature. Graphene grown on copper foil was transferred onto oxidized silicon substrates using an electrochemical delamination method [19]. Poly-methyl methacrylate (PMMA), which acts as a mechanical support layer during the delamination process, was spin-coated onto graphene. After transfer, the samples were placed in a bath of acetone for 6 hours to remove the PMMA and subsequently rapid thermal annealing (RTA) was carried out for 1 hour in argon (95%) / hydrogen (5%) atmosphere at 450°C to minimize PMMA residue. Figure 3b shows an optical micrograph of transferred graphene on a silicon dioxide / silicon (SiO$_2$/Si) substrate. The inset shows large area homogenous graphene transferred onto SiO$_2$/Si substrate.

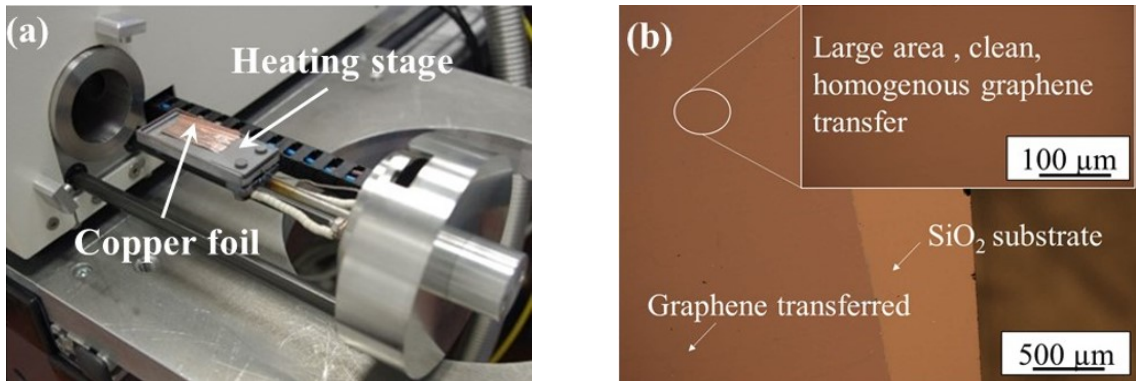

Figure 3: Photograph showing (a) NanoCVD, Moorfield cold wall reactor with Copper foil placed on the heater stage during the loading of the sample, (b) Photograph of large area, clean and homogenous transferred graphene on a SiO$_2$/Si substrate.

Photolithography was carried out to define graphene channels followed by an oxygen (O$_2$) plasma etch of the graphene in the unprotected region. Photoresist was removed by placing the sample in a bath of acetone for 2 hours. Source and drain contacts were defined using image reversal lithography. Various metals were deposited on different samples by evaporation, followed by removal of excessive metal using lift-off processes in suitable solvent. Metals used for the study are gold (Au), nickel (Ni), nickel/gold (Ni/Au), palladium (Pd) and platinum/gold (Pt/Au). A schematic of the entire fabrication process flow is shown in Figure 4.

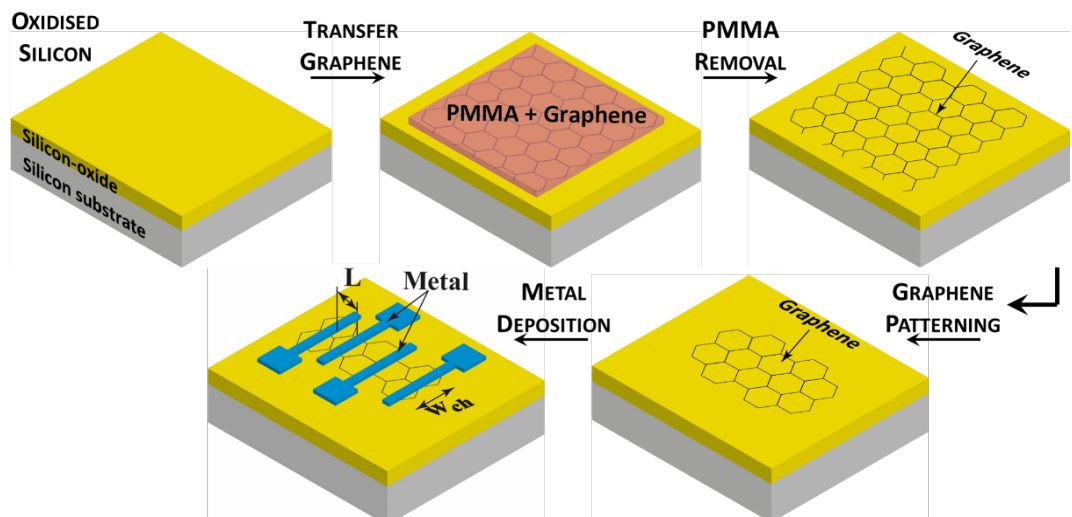

Figure 4: Fabrication process for graphene based back-gated field effect transistor.

The channel widths of the TLM structures varied from 4 μm to 40 μm and the contact spacing (L) was varied from 5 μm to 80 μm, respectively. Figure 5 shows an optical micrograph of one TLM structure with a width of 40 μm. TLM structures were used for the contact resistance determination because the spacing between the contacts (channel length) can be expected to be larger than the carrier mean free path in the channel [5]. TLMs with contact spacing as small as 5 µm were designed to accurately estimate the contact resistance, whereas contact spacing as large as 80 µm were designed to ensure that the measured resistance is dominated by the graphene channel resistance in order to extract the exact sheet resistance ($R_{SH}$) [20].

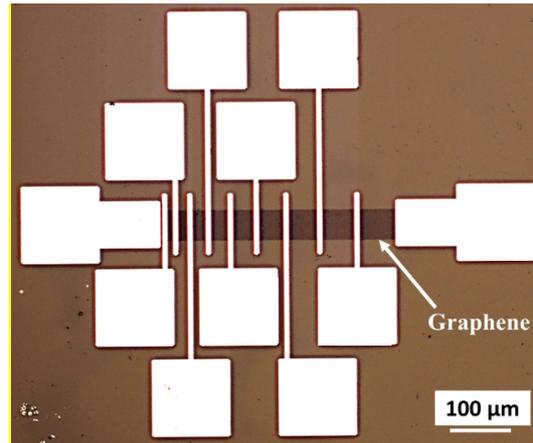

Figure 5: Optical micrograph of a TLM structure with nickel contacts. The channel width of the device is 40 µm.

Electrical measurements were performed with a Karl Süss probe station and a Lakeshore vacuum probe station connected to a Keithley semiconductor analyzer (SCS4200). Back-gate bias was varied from -40 V to +80 V in steps of 0.1 V at fixed drain source voltage ($V_{ds}$) of 50 mV. Measurements were carried out under ambient conditions (21°C and 50% relative humidity) and under vacuum ($10^{-3}$ torr).

RESULTS AND DISCUSSION

AMBIENT CONDITION MEASUREMENTS (21°C AND 50% RELATIVE HUMIDITY):

Twenty devices with channel widths varying from 4 µm to 40 µm were measured. The extracted $R_C$ of various metals to monolayer graphene is plotted in Figure 6. The device width was 40 µm and the measurements were carried out in ambient conditions (21°C and 50% relative humidity). The solid line is the linear fit of the measured data, and the correlation coefficient ($R^2$) lies in the range between 0.9941 – 0.9994. The extracted value of $R_C$ is 500 (±213) Ωµm when gold is used as the metal electrode and is 404(±383) Ωµm for a stack of Ni (25 nm) / Au (50 nm). The value of $R_C$ is 1068 (±515) Ωµm, 2248 (±417) Ωµm and 968(±317) Ωµm when Pt (25 nm) / Au (50 nm), Ni (75 nm) and Pd (75 nm) are used as the metal electrodes, respectively. These results are summarized in Table 1.

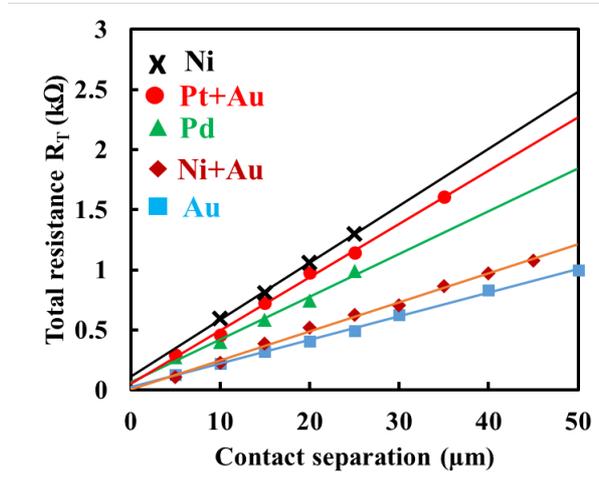

Figure 6: Total resistance as a function of separation between contacts. Metals used to contact graphene are Au (blue rectangle), Ni+Au (brown diamond), Pd (green triangle), Pt+Au (red circles) and Ni (black cross), respectively. The channel width is 40 µm. The solid line is the linear fit of the measured data.

The devices were further measured in a back-gate field effect transistor configuration by applying a back-gate bias to the Si substrate. A fixed drain source voltage ($V_{DS}$) of 50 mV and varying back-gate voltages ($V_G$) from -40 V to +80 V (for devices with gold as the contact metal the applied back-gate was swept from -40 V to +120 V) was applied to the drain-source contacts and substrate, respectively.

Figure 7 shows the transfer characteristics of the back-gated graphene devices with Ni/Au, Pt/Au, Ni and Au as metal electrodes. The value of $V_G$ at which minimum conductance (Dirac voltage) occurs is 10.3 V, -11.7 V, 104.5 V and 14.8 V, for Ni/Au, Pt/Au, Au and Ni electrodes, respectively. The $R_C$ exhibits a clear $V_G$ dependence, is highest at the Dirac point and decreases as the channel is electrostatically doped by the back-gate bias [5]. We note that the devices with Au contacts were fabricated on oxidized silicon wafers with a silicon-dioxide thickness of 325 nm rather than 85 nm. This was necessary to observe the minimum conductivity point, which was above the breakdown voltage of 85 nm thin silicon-dioxide (

Figure 7c). The transfer characteristics exhibit the standard minimum conductivity around $V_{Dirac}$. Moreover, the p and n branches show an asymmetry attributed to $R_C$, which was extrapolated from the transfer characteristics to be larger in the n branch than that in the p branch [5].

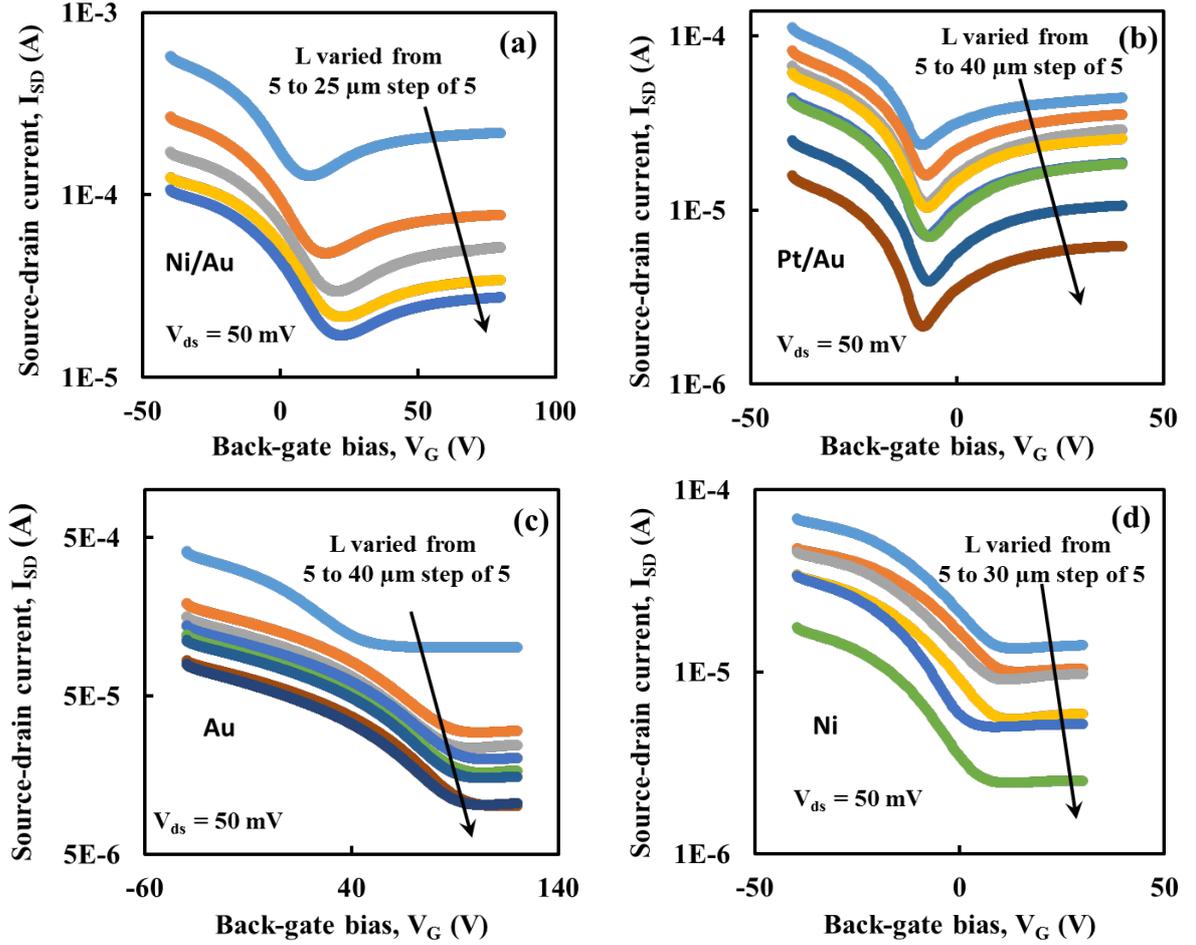

Figure 7: Transfer characteristics $I_{SD} - V_G$ for (a) Ni/Au (b) Pt/Au (c) Au (d) Ni TLM structures.

Contact ($R_C$) and sheet resistance ($R_{SH}$) values extracted from the TLM structures with gold contacts are plotted as a function of back-gate voltage in Figure 8. An $R_C$ of 2100 (±397) Ωμm is extracted at the Dirac voltage of 104.5 V with a correlation coefficient $R^2$=0.998. This value reduces to 92 Ωμm when the channel is heavily p-doped by applying a back gate voltage of $V_G$ = -40 V. The main reason for such low values can be attributed to the attractive interaction between the d-orbitals of gold atoms and the π-orbitals of the $sp^2$ -hybridized carbon, small contact length and point defects [21]. The work function of Au (5.1 eV) is mainly responsible for p doping graphene by extracting electron from it. Back gate bias changes the fermi level in doped graphene thus increasing DOS further reducing $R_C$ [21]. An $R^2$ of 0.9977 is calculated in this case, indicating high fidelity of the data. The extraction of $R_{SH}$ yields a value of 4765 Ω/□ at 104.5 V (Dirac voltage) and 524 Ω/□ at $V_G$ = -40 V. The results are summarized in Table 1. The present results demonstrate that there exists a direct correlation between contact resistivity ($R_CW$) and $R_{SH}$. Contact resistivity ($R_CW$) of nickel-graphene contact is (2248±417) Ωμm and sheet resistance ($R_{SH}$) obtained is 1.919 kΩ/□. Similarly, contact resistivity ($R_CW$) of gold-graphene contact is (500±213) Ωμm and sheet resistance($R_{SH}$) obtained is 0.793 kΩ/□. The results demonstrate that contact resistance depends on the Dirac point energy in graphene and therefore on the charge density in the graphene layer, which is often not controlled in the literature [22]. The length scale at which the Dirac point energy changes is very small (less than 2 nm from the metal surface). Therefore four-probe or transfer lengths cannot incorporate this effect while extrapolating contact resistance [22].The observed correlation between sheet and contact resitance is in line with theoretical predictions [22].

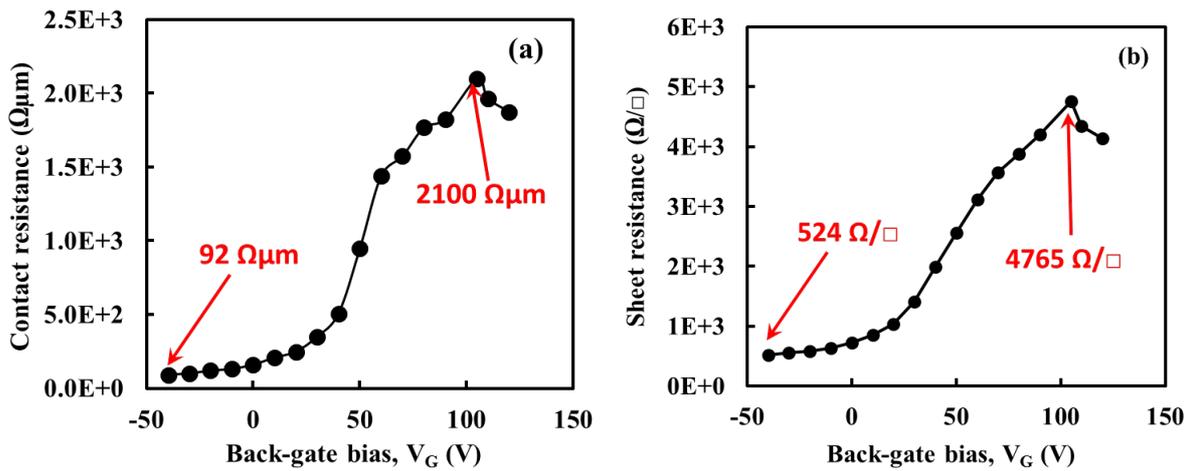

Figure 8: Extraction of (a) contact resistance and (b) sheet resistance at varying back-gate voltages for devices with gold metal electrodes.

MEASUREMENTS UNDER VACUUM ($10^{-3}$ torr)

Back-gated transistors with Ni/Au and Pt/Au contacts were measured under vacuum condition ($10^{-3}$ torr). Figure 9 shows the transfer characteristics of back-gated graphene devices with (a) Ni/Au and (b) Pt/Au, respectively. Measurements under vacuum result in an increase of the $R_C$ value. This increase can be attributed to a shift of the minimum conductance point due to the reduction of random molecular doping/and or humidity [14], which leads to a lower carrier density in the graphene at $V_G = 0$ V. The Dirac point voltage shifts from 10.3 V to 2.5 V and from -11.7 V to -5 V for devices with Ni/Au and Pt/Au contacts, respectively when measurements are carried out under vacuum.

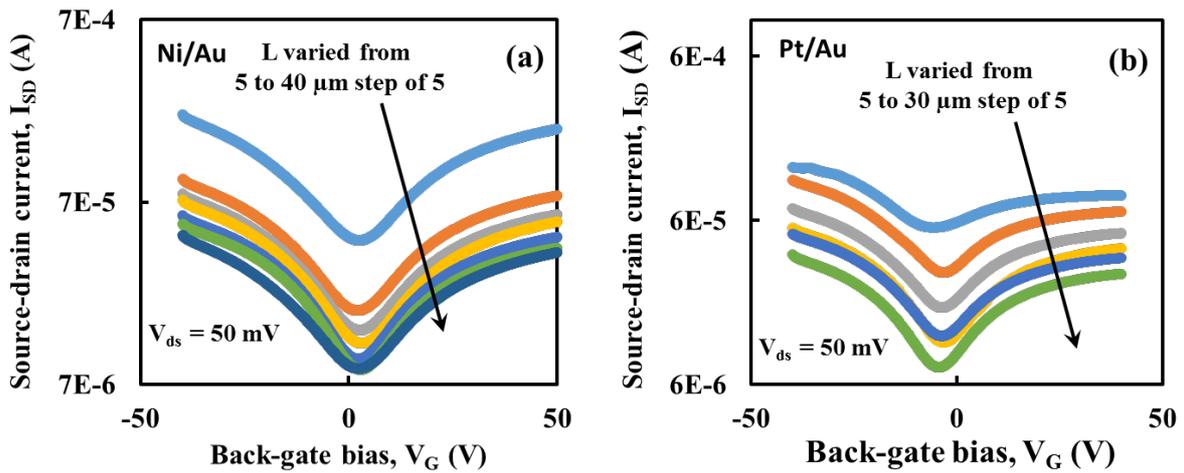

Figure 9: Transfer characteristics $I_{SD}$ -$V_G$ for (a) Ni/Au, (b) Pt/Au TLM structures measured under vacuum.

This result is confirmed in a comparison of the total device resitance $R_T$ as a function of varying L for devices with Ni/Au and Pt/Au metal contacts (Figure 10). Here, the measurements were carried out in ambient conditions (blue squares) and under vacuum (red diamonds) at a back-gate bias of -40 V. For devices with Ni/Au contacts the value of $R_C$ increases from 118 Ωμm to 1600 Ωμm in ambient air and under vacuum, respectively. Similarly, for device with Pt/Au contacts the value of $R_C$ increases from 958 Ωμm to 4768 Ωμm.

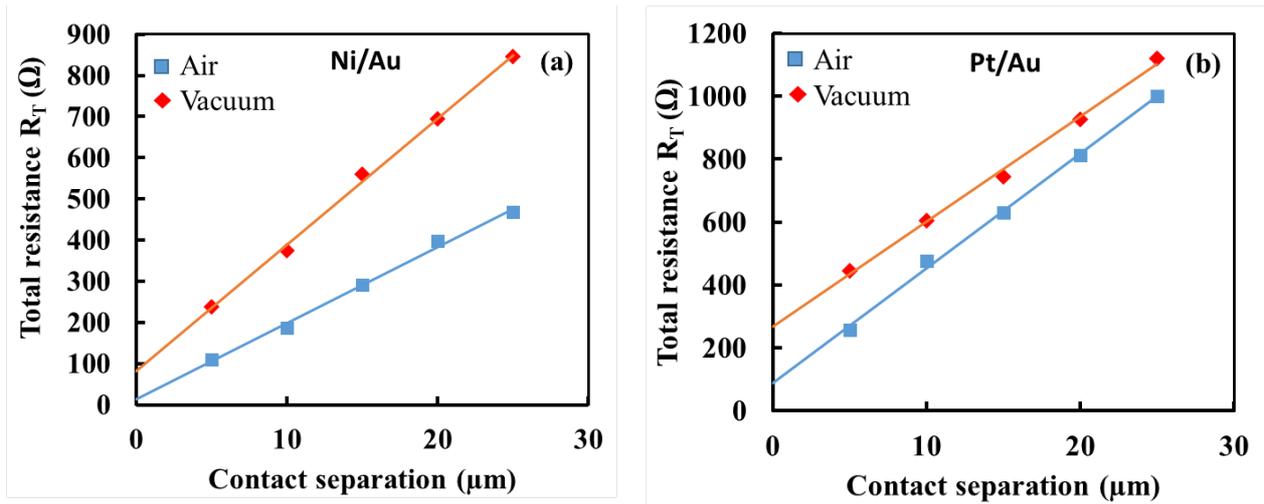

Figure 10: Total resistance as a function of contact separation for (a) Ni/Au and (b) Pt/Au (measurement under vacuum of $10^{-3}$ torr). The channel width is 40 μm. The solid is the linear fit of the measured data.

Effect of Rapid Thermal Annealing (RTA)

A post processing anneal (RTA) was carried out on samples on which Pd and Au were used as metal contacts. RTA was done at 450°C in an atmosphere containing a mixture of Argon (95%) and Hydrogen (5%) under vacuum ($10^{-3}$ torr). Figure 11 shows the total resistance $R_T$ of devices with Pd and Au contacts. This clear reduction in resistance is attributed to an enhancement of the adhesion between metal and graphene. A possible explanation could be that carbon dissolves into the metal during RTA, which could lead to defects in the atomic structure of graphene. This would result in so-called "end-contacts", i.e. covalent bonds between the metal and the graphene, reducing the contact resistance [23].

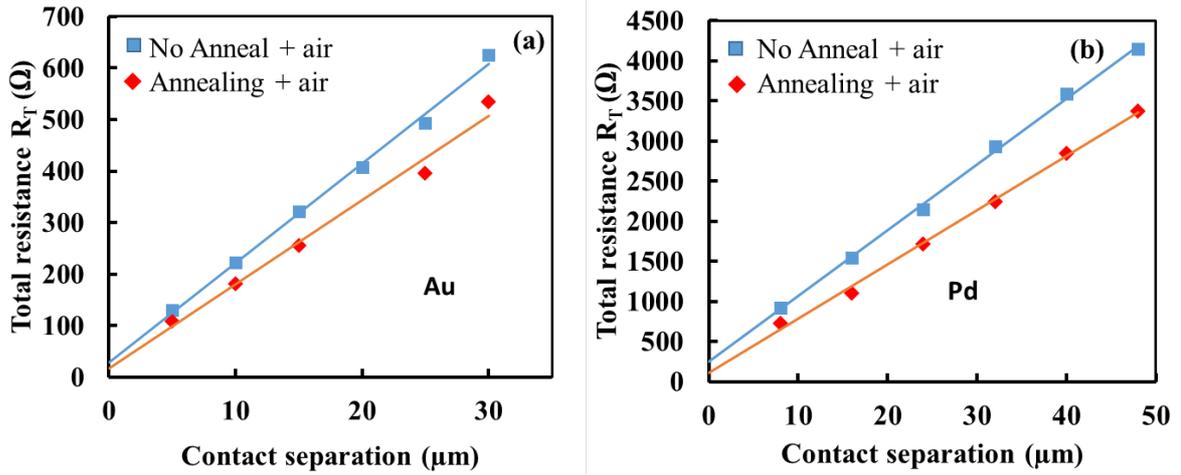

Figure 11: Comparison of total device resistance measured in ambient condition, before and after rapid thermal annealing for (a) Au, channel width in this case is 40 μm and (b) Pd contact, channel width in this case is 17 μm respectively.

The value of contact and sheet resistance of various metal electrodes to monolayer graphene are summarized in Table 1.

Table 1: Contact resistance (Rc) and sheet resistivity ($R_{SH}$) of measured devices with various metal contacts

| Contact metal | Channel width (W) (μm) | $R_{SH}$ (kΩ/□) | $R_CW$ (Ωμm) | Measurement condition | Best reported $R_CW$ (Ωμm) |
|---|---|---|---|---|---|
| Ni (75nm) | 40 | 1.919 | (2248±417) | Air | 100 [24] |
| Au (81nm) | 40 | 0.793 | (500±213) | Air | 340 [21] |
| Pd (75 nm) | 40 | 1.149 | (968±317) | Air | 69 [4] |
| Ni(25nm)+Au(50nm) | 40 | 0.977 | (404±382.9) | Air | - |
| Pt (25nm)+Au(50nm) | 40 | 1.799 | (1068±514.6) | Air | - |

CONCLUSION

A reproducible process to fabricate monolayer graphene based transfer length method structures is presented. Five different metals (gold, nickel, nickel/gold, palladium and platinum/gold) were used to contact monolayer graphene and the contact of these metals to graphene is compared and presented. In addition, the sheet resistance is extracted. The doping concentration and the broadening of the density of states in the graphene under the metal and in the graphene forming the channel are different, leading to different sheet resistances [5]. The lowest contact resistance of 92 Ωμm was achieved for devices with gold contacts, extracted at a back-gate bias of -40 V. We also report the dependence of the contact resistance on the applied back-gate bias. The contact resistance values increase when measurements are carried out under vacuum as compared to ambient conditions (21°C and 50% relative humidity).Finally, post processing annealing reduces the contact resistance. In summary, we provide an extensive experimental investigation of the contact properties of various metals to graphene, including the influence of processing and measurement conditions. The procedures lead to excellent contact resistance values of less than 100 Ωμm.

ACKNOWLEDGEMENTS

The authors' would like to thank the support from the European Commission through an ERC starting grant (InteGraDe, 307311), an FP7 project (GRADE, 317839), the German Research Foundation (DFG, LE 2440/1-1) and the German BMBF (NanoGraM, 03XP0006C).ACKNOWLEDGEMENTS

The authors' would like to thank the support from the European Commission through an ERC starting grant (InteGraDe, 307311), an FP7 project (GRADE, 317839), the German Research Foundation (DFG, LE 2440/1-1) and the German BMBF (NanoGraM, 03XP0006C).